\documentclass[showpacs,aps,pre,floatfix,amsmath,amssymb]{revtex4}
\usepackage{graphicx}
\usepackage{pstricks}

\begin{document}

\title{Low temperature Glauber dynamics under weak competing interactions}

\author{M. D. Grynberg} 

\affiliation{Departamento de F\'{\i}sica, Universidad Nacional de  
La Plata, (1900) La Plata, Argentina}

\begin{abstract}
We consider the low but nonzero temperature regimes of the Glauber dynamics 
in a chain of Ising spins with first and second neighbor interactions $J_1,\, J_2$. 
For $0 < -J_2 / \vert J_1 \vert < 1$ it is known that at $T = 0$ the dynamics is both 
metastable and non-coarsening, while being always ergodic and coarsening in the 
limit of $T \to 0^+$. Based on finite-size scaling analyses of relaxation times, here 
we argue that in that latter situation the asymptotic kinetics of small or weakly 
frustrated $-J_2/ \vert J_1 \vert$ ratios is characterized by an almost ballistic dynamic 
exponent $z \simeq 1.03(2)$ and arbitrarily slow velocities of growth. By contrast, for 
non-competing interactions the coarsening length scales are estimated to be almost 
diffusive.
\end{abstract}

	
\maketitle

\section{Introduction}

Processes usually referred to as phase ordering dynamics, or coarsening, or domain 
growth following an instantaneous quench from a high to a sub-critical temperature 
ordered phase, have been the focus of intense research efforts over the past few 
decades \cite{Puri}. At the same time, many interesting issues continue to arise in 
the field. An example of this is posed by the zero temperature Glauber dynamics 
\cite{Glauber,Felderhof,Spirin} of Ising models in which energy-increasing spin flips 
are rejected. As a result, in higher dimensions the dynamics is not always able to 
reach its ground states but rather probably gets trapped in metastable situations 
\cite{Spirin}. This lack of ergodicity may even appear in one-dimension when 
competing interactions are considered \cite{Redner}. In particular, this is the case 
of the 1D axial next-nearest-neighbor Ising model \cite{Tanaka,Liebmann} of spins 
$S = \pm 1$, and energy configurations  
\begin{equation}
\label{energy}
E = - J_1 \sum_i S_i \, S_{i+1} - J_2 \sum_i S_i \, S_{i+2}\,,
\end{equation}
\vskip -0.07cm
\noindent
in which frustration arises when combining antiferromagnetic (AF) $J_2$-\,couplings 
($J_2 < 0$) with $J_1$-\,exchanges of any sign \cite{note1}. From its original 
formulation in higher dimensions to account for (equilibrium) structure factors in 
rare-earths along their single competing axes \cite{Elliot}, this model has been  
recurrently investigated ever since \cite{Liebmann,Selke,Rastelli}. When it comes to 
the dynamics, the magnetic relaxation of recently synthesized molecular chains with 
strong Ising anisotropy \cite{exp} was also considered in this framework by means of a 
decoupling approximation of the underlying master equation \cite{Pini}. In contrast to 
previous decoupling schemes \cite{Yang} and numerical simulations \cite{Sen}, it was 
found that in the strongly frustrated but ergodic regime $- J_2 > \vert J_1 \vert$, the 
Glauber relaxation time $\tau$ becomes of the order of the equilibrium correlation 
length $\xi$ \cite{Pini}. Since at evolution times comparable with $\tau$ the average 
domain size is $\propto \xi$, this suggests that the ordering scale should coarsen 
ballistically (linearly in time), rather than in a diffusive \cite{Yang} or subdiffusive 
manner \cite{Sen}. 

Here we will not venture into that latter controversy but focus instead on the less 
studied situation of weakly competing ratios $r \equiv -J_2 /\vert J_1 \vert$ in the limit  
of $T \to 0^+$. In marked contrast to the usual Glauber dynamics \cite{Glauber, 
Felderhof} ($J_2 = 0$) and irrespective of how small the frustration might be, note that 
at $T = 0$ the length scale of magnetic domains does {\it not} coarsen \cite{Redner}. 
This is because for $0 < r < 1/2$ (say with $J_1 >  0$ and starting from a disordered 
phase), the ferromagnetic (F) state is unreachable as kinks or domain walls cannot 
approach each other less than two lattice spacings, thus precluding their annihilation 
via single-spin flips \cite{Redner}. In turn, when $1/2 < r < 1$ there is no way to reach 
the four-fold degenerate ground state (consisting of consecutive pairs of oppositely 
oriented spins $\cdots \bullet \bullet \circ \circ \cdots$), because the creation of kink 
pairs is forbidden at $T\!=0$ and so isolated  domains of width $\ge 3$ cannot break 
up \cite{Redner}. (These issues get clear with the aid of Table~\ref{tab1}, introduced 
later on in Sec. II). In either case, the system rapidly ends up fluctuating in a metastable 
phase of domains of length $\ge 2$, in which kinks can wander around at no energy 
cost while keeping their density fixed \cite{Redner,Sen}. In the thermodynamic limit 
this latter approaches a finite fraction $\rho=(5 - \sqrt 5)/10$\, \cite{Sen}, so large 
domain lengths $\lambda$ are distributed as $\rho\, e^{-\rho \lambda}$ \cite{Feller} 
and the average domain size cannot but remain bounded as $\sim \int_0^{\infty} 
\lambda\,\rho\, e^{-\rho \lambda}\,d \lambda = 1/\rho$. As for $r = 1/2$, there the 
ground state is accessible but highly degenerate, and its order range also turns out to 
be finite, i.e. $\xi = 1/\ln g$\, with golden ratio $g = \frac{1}{2}(\sqrt 5+ 1)$
\cite{Tanaka,Liebmann}.

Despite this non-coarsening scenario, in what follows we will consider the large time 
scaling effects introduced by a small but nonzero temperature in this weakly frustrated 
dynamics. Although at $T=0$ all AF strings disappear exponentially fast in time 
\cite{Redner}, and the local persistence (i.e., the probability for a spin to remain in its 
original state at a given time) decreases as a stretched exponential form \cite{Sen},
yet at infinitesimal temperatures the actual relaxation time gets arbitrarily large, just
because of the energy barriers responsible of the mentioned metastable phase. In 
addition, once ergodicity is restored note that for $ 0 < r < 1/2$ and temperatures 
much below the so-called disorder line:  $\cosh K_1 = e^{2 K_2}$ \cite{Tanaka, 
Liebmann,Selke}, the now accessible equilibrium correlation length grows unbounded 
as $\sim\frac{1}{2}\,e^{(2K_1 + 4K_2)}$, while for $r >1/2$ it does so as $\sim 2\, 
e^{\vert K_1 + \,2K_2 \vert}$ \cite{note2}, where $K_i\equiv J_i/k_B T$. Thus, following 
critical dynamic theories \cite{Hohenberg}, in approaching $T=0^+$ both of these 
growing space and time scales should behave as $\tau \propto \xi^z$ with a dynamic 
exponent $z$ characteristic of the universality class to which the dynamics belongs. In 
the long time limit it basically describes how fast the coarsening scale is spreading, and 
in a way that if spatial coordinates are rescaled as $x \to x/t^{1/z}$ the probability 
distribution of domain lengths remains invariant \cite{Puri}. On the other hand, in 
nearing the critical regime the correlation length becomes comparable to the system 
size $L$. Hence it is customary to think that the above scaling relation can be traded in 
practice for the finite-size behavior $\tau_{\!_L} \propto L^z$, provided $L$ is taken 
sufficiently large \cite{Henkel}. Thus, in the following sections we attempt to provide 
an estimation of this exponent from the information of finite systems. To that aim, 
firstly we will recast the master equation \cite {Kampen} governing these Markov 
processes in terms of a quantum spin representation of the associated Glauber 
operator, i.e. the Liouvillian or evolution matrix of the problem \cite {Kampen}. This 
latter lends itself more readily for a finite-size scaling analysis of actual relaxation times 
as these are embodied in spectral gaps which we will subsequently evaluate by exact 
diagonalizations \cite{Lanczos}.

The layout of this work is organized as follows. Sec.\,II outlines the basic transition 
probability rates of these processes (all nonzero at $T > 0$) using a kink or dual 
representation here thought of as a spin-$\frac{1}{2}$ dynamics. Exploiting detailed 
balance \cite{Kampen} the associated evolution operator can then be brought to a 
symmetric representation via simple nonunitary spin transformations. In part, this 
simplifies the numerical analysis of Sec. III in which spectrum gaps are obtained via 
standard recursive techniques \cite{Lanczos} in various situations, both for competing 
and non-competing interactions. This provides a sequence of finite-size approximants 
to dynamic exponents which are then combined with extrapolations \cite{Henkel,
Guttmann}, so as to partly remedy unavoidable size effects. We close with Sec.\,IV 
which contains a recapitulation along with brief remarks on open issues and possible 
extensions of this work.

\section{Stochastic Dynamics}

As is known \cite{Glauber,Felderhof}, in the Glauber dynamics the heat bath is 
thought of as inducing fluctuations in the form of single-spin flip processes, thus 
causing the states $\vert S\, \rangle  = \vert S_1,\,\dots\,, S_L \rangle$ to change 
randomly. For instantaneous quenches the transition probability rates  per unit time 
$W ( S \to  S')$ between two configurations $\vert S \rangle, \vert S'\rangle$ (here 
differing in the state of one spin), are taken time independent and chosen to satisfy 
the detailed balance condition \cite{Kampen}
\begin{equation}
\label{DB}
P_B (S) \,W(S \to S') = P_B (S') \, W(S' \to S), \;\forall \; \vert S \rangle,\vert S'\rangle\,,
\end{equation}
so as to enforce the system to relax towards the Boltzmann distribution $P_B (S) 
\propto  e^{- \beta E(S)}$ at large times \cite{Kampen}. Hereafter  temperatures 
are measured in energy units; or equivalently, the Boltzmann constant in $\beta 
\equiv 1/(k_B T)$ is set to 1. For single-spin flip processes the simplest choice of rates 
complying with Eq.\,(\ref{DB}) corresponds to that of Glauber \cite{Puri,Glauber, 
Felderhof}, which adapted to Eq.\,(\ref{energy}) reads
\begin{equation}
\label{rates}
W ( S_i \to  - S_i)  = \frac{\alpha}{2}\, \big\{\,1\, - \, S_i\, \tanh \big[\, K_1 \left(\, 
S_{i-1} + S_{i+1} ,\right) + K_2 \left(\,S_{i-2} + S_{i+2}\,\right) \big]\,\big\}\,.
\end{equation}
Here $\alpha^{-1}$ simply sets the time scale of the microscopic process, and 
from now on is taken equal to unity. Just as frustration may occur irrespective of 
nearest-neighbor (NN) interactions being F or AF \cite{note1}, note that the dynamics 
is not affected either by the nature of those couplings. To check this out consider for 
instance the mapping $S_{2i} \to -S_{2i}$, while leaving the odd spins unchanged.
Assuming both periodic boundary conditions (PBC) and an even number of spins 
-henceforth considered throughout- then $S_i S_{i+1}\!\to -S_i S_{i+1}$ and $S_i 
S_{i+2} \to \!S_i S_{i+2}.\!$ Therefore, by taking $J_1 \!\to -J_1$ the rates of 
Eq.\,(\ref{rates}) are left invariant and so is the mapped dynamics because the 
single-spin flip process maps onto itself. (This reasoning however would not apply 
to a dynamics such as the Kawasaki one \cite{Puri} as the mapping would then allow 
parallel NN spins to flip). Thus, without lose of generality in what follows we shall 
consider, say, ferromagnetic NN interactions along with $J_2$ couplings of either sign. 
We shall get back to this point by the end of this Section.

As is schematized in Table~\ref{tab1} there are basically eight processes (some 
of them symmetric by reflection in the  active spin), which for next convenience we 
now disaggregate in two sets of dual events namely, pairing and diffusion of kinks.  
Also, after introducing the parameters 
\begin{equation}
\label{pq}
P = 2\, (K_1 + K_2),\,\;  Q = 2\, (K_1 - K_2),
\end{equation}
there we list the transition rates of these processes along with their energy  
gradients. Owing to these latter note in passing that at $T=0$ and  $0 < r < 1$ all AF 
domains would be irreversibly removed by pairing, while the remaining set of diffusion 
processes would maintain all kinks separated in at least two lattice spacings. (See
related discussion of Sec.\,III\,A). This is in line with the metastable picture given 
earlier on in Sec.\,I, as it should, though that is to be contrasted with the actual 
non-zero low temperature dynamics and towards which we next turn. 
\begin{table}[htbp]
\vskip 0.3cm
\begin{center}
\begin{tabular} {c  c  c  c  c}
 \hline \hline
Pairing process  &   \hskip 0.75cm Rate $( \rightleftarrows)$ & \hskip 0.7cm  
$\beta \Delta E\, ( \rightleftarrows)$ & \hskip 0.7cm S- element
& \hskip 0.5cm Projector
\vspace{0.05cm}
\\  
 \hline \hline
\vspace{-0.3cm} \\


$\circ \;\, \circ\, {\blue \vert \bullet  \vert} \circ\; \circ \;\;\;
\rightleftarrows \;\;\, \circ\;\;  \circ\;\; {\blue \circ}\;\; \circ\;\; \circ$  
& \hskip 0.75cm  $\frac{1}{2} (1\pm \tanh P)$
&  \hskip 0.75cm  $\mp 2 P$ 
&  \hskip 0.75cm  $\frac{1}{2}\,{\rm sech}\,P$ 
&  \hskip 0.5cm ${\cal P}^{(1)}$
\vspace{0.3cm}\\

$\bullet\; \vert \circ {\blue \vert  \bullet  \vert}  \circ\; \circ \;\;\;
\rightleftarrows \;\;\,  \bullet\; \vert \circ\;\; {\blue \circ}\;\; \circ\;\;\circ$  
& \hskip 0.75cm   $\frac{1}{2} (1\pm \tanh 2 K_{_{\!1}})$
&  \hskip 0.75cm  $\mp 4 K_{_{\!1}}$
&  \hskip 0.75cm  $\frac{1}{2}\,{\rm sech}\,2 K_{_{\!1}}$
&  \hskip 0.5cm ${\cal P}^{(2)}$
\vspace{0.3cm}\\

$\circ \;\, \circ\, {\blue \vert \bullet \vert} \circ \vert\; \bullet \;\;\,
\rightleftarrows \;\;  \circ\;\; \circ\;\; {\blue \circ}\;\; \circ \vert\; \bullet$  
& \hskip 0.75cm   $\frac{1}{2} (1\pm \tanh 2 K_{_{\!1}})$
&  \hskip 0.75cm  $\mp 4 K_{_{\!1}}$
&  \hskip 0.75cm  $\frac{1}{2}\,{\rm sech}\,2 K_{_{\!1}}$
&  \hskip 0.5cm ${\cal P}^{(3)}$
\vspace{0.3cm}\\

$\bullet\; \vert \circ {\blue \vert \bullet \vert} \circ \vert\; \bullet \;\;\,
\rightleftarrows \;\;  \bullet\; \vert \circ\;\; {\blue \circ}\;\; \circ  \vert \;\bullet$  
& \hskip 0.75cm   $\frac{1}{2} (1\pm \tanh Q)$
&  \hskip 0.75cm  $\mp 2 Q$  
&  \hskip 0.75cm  $\frac{1}{2}\,{\rm sech}\,Q$
&  \hskip 0.5cm ${\cal P}^{(4)}$
\vspace{0.3cm}\\


\hline \hline
Diffusion process  &   \hskip 0.75cm Rate $( \rightleftarrows)$ & \hskip 0.7cm  
$\beta \Delta E\, ( \rightleftarrows)$ & \hskip 0.7cm S- element
& \hskip 0.5cm Projector
\vspace{0.05cm}
\\   \hline \hline
\vspace{-0.3cm} \\

$\;\,\bullet\;\, \bullet\;\, {\blue  \bullet \; \vert} \circ\;  \circ  \;\;\,
\rightleftarrows  \;\; \bullet\;\,\bullet\,{\blue \vert \; \circ} \;\, \circ\;\, \circ$  
& \hskip 0.75cm $ 1/2 $ 
&  \hskip 0.75cm  $ 0 $
&  \hskip 0.75cm  $ 1/2 $
&  \hskip 0.5cm ${\cal P}^{(1)}$
\vspace{0.3cm}\\

$\,\circ\; \vert \bullet\;\, {\blue \bullet \; \vert} \circ\; \circ  \;\;\,
\rightleftarrows  \;\; \circ\; \vert \bullet {\blue \vert\;  \circ} \;\, \circ\;\, \circ$  
& \hskip 0.75cm   $\frac{1}{2} (1\pm \tanh 2 K_{_{\!2}})$
&  \hskip 0.75cm  $\mp 4 K_{_{\!2}}$
&  \hskip 0.75cm  $\frac{1}{2}\,{\rm sech}\,2 K_{_{\!2}}$
&  \hskip 0.5cm ${\cal P}^{(2)}$
\vspace{0.3cm}\\

$\,\circ\;\, \circ \,{\blue \vert \;  \bullet} \;\,  \bullet \vert\; \circ  \;\;\,
\rightleftarrows  \;\; \circ\;\, \circ\;\,  {\blue \circ\;  \vert} \bullet \vert\, \circ$  
& \hskip 0.75cm   $\frac{1}{2} (1\pm \tanh 2 K_{_{\!2}})$
&  \hskip 0.75cm  $\mp 4 K_{_{\!2}}$
&  \hskip 0.75cm  $\frac{1}{2}\,{\rm sech}\,2 K_{_{\!2}}$
&  \hskip 0.5cm ${\cal P}^{(3)}$
\vspace{0.3cm}\\

$\,\bullet\; \vert \circ \, {\blue \vert\;  \bullet} \;\, \bullet \vert \, \circ\;\;\,
\rightleftarrows  \;\; \bullet\; \vert \circ\; {\blue \circ\; \vert} \bullet \vert\, \circ$  
& \hskip 0.75cm $ 1/2 $ 
&  \hskip 0.75cm  $ 0 $
&  \hskip 0.75cm  $ 1/2 $
&  \hskip 0.5cm ${\cal P}^{(4)}$
\vspace{0.3cm}\\

\hline \hline

\end{tabular}
\end{center}
\vskip 0.35cm
\caption{(Color online) Glauber transition probabilities, energy changes, and 
symmetrized (S) non-diagonal matrix elements of the evolution operator transformed 
as in Eq.\,(\ref{s-rates}), for both kink pairing and diffusion processes under $J_1$ 
and $J_2$ interactions. Filled and empty circles denote original spins with opposite 
orientations in turn conforming kinks (vertical lines) on the dual chain. Upper and 
lower signs stand respectively for the forward ($\rightarrow$) and backward
($\leftarrow$) processes brought about by flipping central spins. All events are
classified according to the projector types defined in Eq.\,(\ref{proj}).}
\label{tab1}
\end{table}

\subsection{Evolution operators}

The transition rates referred to in Eq.\,(\ref{rates}) form part of the so-called master 
equation \cite{Kampen} controlling the time evolution of these Markovian processes. 
It determines the probability distribution $\vert P(t) \,\rangle \equiv \sum_S P (S,t)\,
\vert S\, \rangle\,$  to observe the system in one the above states $\vert S \rangle$ 
at a given time $t$. Following some of the formal analogies between stochastic 
and quantum systems \cite{Gunter} we now think of the master equation as a
Schr\"odinger one in an imaginary time, that is
\begin{equation}
\label{master}
\frac{\partial}{\partial t}\, \vert P (t)\, \rangle = - H\, \vert P (t) \,\rangle\,,
\end{equation}
thus allowing by formal integration to obtain the probability distribution at
subsequent moments from the action of $H$ on a given initial condition, i.e. 
$\vert P (t) \,\rangle = e^{- Ht}\,\vert P (0) \,\rangle$. Here, the `Hamiltonian'
(not necessarily hermitian in this context) or Liouvillian operator generating 
the dynamics is defined through the matrix elements
\begin{equation}
\langle\,S'\,\vert\,H \,\vert\,S\,\rangle = 
\label{elements}
\begin{cases}
- W (S \to S'),\;\;\;\;\;\;\;\;\;\;\;{\rm for}\;\;S \ne S', 
\vspace{0.1cm} \cr
\sum_{S'\ne S}\, W (S \to S'),\;\;{\rm for}\;\;S = S',
\end{cases}
\end{equation}
which, by conservation of probability, constrain all columns to add up to zero. Thereby it 
can be shown \cite{Kampen,Gradshteyn} that the steady state corresponds to a unique 
$H$-eigenmode with eigenvalue $\lambda_0 = 0$, whereas the relaxation time of any 
observable average $\sum_S {\cal O} (S) P (S,t)$ (such as the magnetization or 
equal-time pair correlations), is upper bounded  by $1/{\rm Re}\,( \lambda_1) > 0$, 
with $\lambda_1$ being the first excitation level of the $H$-spectrum.

Since the phase space dimension of these operators grows exponentially with 
the system size, an operational counterpart of Eq.\,(\ref{elements}) is needed
to implement the recurrence techniques of Sec.\,III. On the other hand, to downsize 
at least part of those memory requirements, we now turn to a two-to-one mapping 
$\sigma_i \equiv - S_i S_{i+1}$ in which new Ising variables standing on dual 
locations denote the presence/absence of the kinks referred to above. As depicted 
in Table~\ref{tab1}, in that representation the Liouvillian must account for both of 
the process types brought about by flipping one of the originals $S_i$. In that regard, 
if we think of the states $\vert \sigma_1,\,\dots\,\sigma_L \rangle$ as representing 
configurations of $\frac{1}{2}$-spinors (say in the $z$ direction), then the
operational analog of Eq.\,(\ref{elements}) can be readily obtained with the aid of 
raising and lowering operators $\sigma^+,\sigma^-$. Clearly, the non-diagonal parts 
of $H$ should involve terms $\propto\, \sigma^{\pm}_{i-1} \sigma^{\pm}_i$ as 
well as other ones $\propto\, \sigma^{\pm}_{i-1} \sigma^{\mp}_i$ associated 
respectively to the pairing and diffusion events sketched above (say on dual locations 
$i-1,\,i$). However, note that those processes should be weighted by rates which 
actually depend on the kink occupation $\hat n = \sigma^+ \sigma^- = \frac{1}{2} 
(1+\sigma^z)$ and vacancy $\hat{\rm v} = 1 - \hat n$ numbers of the nearest locations 
surrounding the active ones (see Table~\ref{tab1}). Thus, to reproduce the correlations 
associated to of each of these events, here we classify them according to projectors 
defined as
\begin{subequations}
\label{proj}
\begin{eqnarray}
{\cal \hat P}_i^{(1)}\!\! &=&  \! \hat {\rm v}_{i-2}\, \hat {\rm v}_{i+1}\,,\;\;\;
{\cal \hat P}_i^{(2)}=\,\hat n_{i-2}\,\hat {\rm v}_{i+1}\,,\\
{\cal \hat P}_i^{(3)}\!\!  &=& \! \hat {\rm v}_{i-2}\; \hat n_{i+1}\,,\;\;\,
{\cal \hat P}_i^{(4)}=\,\hat n_{i-2}\; \hat n_{i+1}\,,
\end{eqnarray}
\end{subequations}
to which we assign in turn the variables $\{x_1, \, x_2, \, x_3, \, x_4\}\equiv \{P, 2K_1, 
2K_1, Q\}$, and  $\{y_1, \, y_2, \, y_3,\, y_4\} \equiv \{0, 2 K_2, -2 K_2, 0\}$. Then, 
it is straightforward to verify that the non-diagonal ($nd$) parts of the evolution 
operator accounting for those correlated pairing and diffusion rates are each given by
\vspace{0.1cm}
\begin{subequations}
\begin{eqnarray}
\label{pairing}
H_{nd}^{^{(pair)}}&=& - \sum_i\, \sum_j {\cal \hat P}_i^{(j)}\! \left[\,f (x_j)\,
\sigma^-_{i-1}\,\sigma^-_i + f (-x_j)\,\sigma^+_i\,\sigma^+_{i-1} \,\right]\,,\\
\label{diffusion}
H_{nd}^{^{(di\!f\!f)}} &=& - \sum_i\, \sum_j {\cal \hat P}_i^{(j)}\! \left[\,
f (y_j)\,\sigma^+_{i-1}\,\sigma^-_i + f (-y_j)\,\sigma^+_i\,\sigma^-_{i-1}\,\right]\,,
\end{eqnarray}
\end{subequations}
where $f (u) \equiv \frac{1}{2} (1 + \tanh u)$. 

Now we can exploit detailed balance to readily bring these expressions into symmetric 
operators. For that purpose it suffices to consider the kink energies $E_{\sigma} =
\sum_i \left( J_1\,\sigma_i -J_2\,\sigma_i\, \sigma_{i+1} \right)$ associated to 
Eq.\,(\ref{energy}) with which we construct the diagonal non-unitary similarity 
transformation $\mathbb D \,\vert \sigma\rangle\equiv e^{\frac{\beta}{2}\, 
E_{\sigma}} \vert \sigma \rangle$. Therefore under this latter, the generic 
non-diagonal elements of Eq.\,(\ref{elements}) transform as
\vskip -0.5cm
\begin{equation}
\label{s-rates}
W ( \sigma \to \sigma') \to e^{\,{\frac{\beta}{2}}\, \left(E_{\sigma'} - E_{\sigma}
\right)}\, W ( \sigma \to \sigma')\,.
\end{equation}
Since in the kink representation the $\sigma$-\,rates also comply with the detailed 
balance condition (\ref{DB}), then clearly these transformed non-diagonal elements 
become symmetric under $\mathbb D$. Identifying this diagonal operator simply with 
$\exp\left[\,\frac{1}{2}\sum_i \left(K_1\,\sigma^z_i - K_2\,\sigma^z_i\sigma ^z_{i+1}
\right) \right]$, then the above pairing and hopping terms will transform respectively as
\begin{subequations}
\begin{eqnarray}
\phantom{\Big[}
 \sigma^{\pm}_{i-1} \,\sigma^{\pm}_i  & \to & 
\exp \left[ \mp K_2 \left( \sigma^z_{i-2} + \,\sigma^z_{i+1}\right) 
\pm 2 K_1 \right]\; \sigma^{\pm}_{i-1} \,\sigma^{\pm}_i\;,\\
\phantom{\Big[}
\sigma^{\pm}_{i-1} \,\sigma^{\mp}_i  & \to &
\exp \left[ \pm K_2 \left( \sigma^z_{i-2} - \sigma^z_{i+1} \right)\right]\,
\sigma^{\pm}_{i-1} \,\sigma^{\mp}_i\;,
\end{eqnarray}
\end{subequations}
(compare with symmetrized elements of  Table~\ref{tab1}), whilst the diagonal 
projectors of Eq.\,(\ref{proj}) remain unchanged. Thus, after some algebraic 
manipulations we are finally led to the symmetric counterparts of Eqs.\,(\ref{pairing}) 
and (\ref{diffusion}), namely
\begin{subequations}
\begin{eqnarray}
\label{s-pairing}
{\cal H}_{nd}^{^{(pair)}} & = & - \frac{1}{4}\,\sum_i \left[\, A_{_+} + 
B \left(\, \sigma^z_{i-2} + 
\sigma^z_{i+1}\right) +  A_{_-}\,\sigma^z_{i-2} \, \sigma^z_{i+1} \,\right]
\left(\,\sigma^+_{i-1}\,\sigma^+_i + {\rm h.c.}\,\right)\,,\\
\label{s-diffusion}
{\cal H}_{nd}^{^{(di\!f\!f)}} & = & - \frac{1}{4}\,\left(\,1 + {\rm sech}\, 2K_2 \,
\right)\,\sum_i  \left(\,1+\tanh^2\!K_2\; \sigma^z_{i-2}\, \sigma^z_{i+1}\,\right)
\left(\,\sigma^+_{i-1}\,\sigma^-_i + {\rm h.c.}\,\right)\,,
\end{eqnarray}
\end{subequations}
where the $A_{\pm}$ and $B$ coefficients are given by
\begin{subequations}
\label{coef1}
\begin{eqnarray}
A_{\pm} &=& \frac{1}{2} \left(\,{\rm sech}\,Q + {\rm sech}\,P\,\right)
 \pm {\rm sech}\, 2K_1 \;,\\
B &=&  \frac{1}{2} \left(\,{\rm sech}\,Q - {\rm sech}\,P\,\right) \,.
\end{eqnarray}
\end{subequations}

When it comes to the  diagonal terms of Eq.(\ref{elements}), in the kink representation 
these basically count the number of ways in which a given configuration can access to 
other ones either by NN pairing or hopping. To probe and weight these attempts with 
the rates of Table~\ref{tab1} we resort once more to the projectors and number 
operators referred to in Eq.\,(\ref{proj}), and in terms of which the diagonal ($d$) 
parts associated to those two type of processes can be summarized each as
\begin{subequations}
\begin{eqnarray}
H_d^{^{(pair)}}\!\!\!\! &=& \sum_i\, \sum_j {\cal \hat P}_i^{(j)}\! \left[\,
f (x_j)\, \hat n_{i-1}\,\hat n_i + f (-x_j)\,\hat{\rm v}_{i-1}\, \hat{\rm v}_i \,\right]\,,\\
H_d^{^{(diff)}}\!\!\!\! &=& \sum_i\, \sum_j {\cal \hat P}_i^{(j)}\! \left[\,
f (y_j)\, \hat{\rm v}_{i-1}\, \hat n_i +  f (-y_j)\,\hat n_{i-1}\, \hat{\rm v}_i\,\right]\,.
\end{eqnarray}
\end{subequations}
Evidently by construction ${\cal H}_d \equiv H_d^{^{(pair)}}\!\!\! + 
H_d^{^{(diff)}}\!$ remains invariant under $\mathbb D\,$, thus collecting all 
projector contributions and defining  further coefficients
\vskip -0.5cm
\begin{subequations}
\label{coef2}
\begin{eqnarray}
C_{\pm} &=& \frac{1}{2} \left(\,\tanh\,Q + \tanh\,P\,\right) \pm \tanh\, 2K_1 \;,\\
D_{\pm} &=&  \frac{1}{2} \left(\,\tanh\,Q - \tanh\,P\,\right) \pm \tanh\, 2K_2 \,,
\end{eqnarray}
\end{subequations}
we are ultimately left with diagonal terms in a uniform field, along with two- and 
three-body interactions of the form
\begin{equation}
\label{diagonal}
{\cal H}_d = \frac{1}{4}\,\sum_i \left[ \left(\, C_{_{\!+}} + D_{_{\!-}}\, 
\sigma^z_{i+1} +  D_{_{\!+}}\,\sigma^z_{i+2}\right) \sigma^z_i 
+ \frac{1}{2}\, C_{_{\!-}} \! \left(\sigma^z_{i-1}\! + \sigma^z_i \right)
\sigma^z_{i-2} \,\sigma^z_{i+1} \right] + \frac{L}{2}\,.
\end{equation}

Together with Eqs.\,(\ref{s-pairing}) and (\ref{s-diffusion}), this latter result completes 
the construction of the operational analog of Eq.\,(\ref{elements}). It is worth noting 
that for $J_2 = 0$  all many-body couplings disappear ($C_{_{\!-}}\! = D_{\pm} =0$)  
along with both corrrelated pairing and hopping terms ($A_{_-}\! = B =0$), so ${\cal 
H}_d + {\cal H}_{nd}^{^{(pair)}}\!\!\!+{\cal H}_{nd}^{^{(di\!f\!f)}}\!\!$ fully 
recovers the bilinear form of Ref.\,\cite{Felderhof}. In that case the evolution operator 
can be diagonalized exactly, and its spectrum reduces to the $2^{L-1}$ manners of 
filling a band of elementary fermionic excitations\, $E_q = 1 - \tanh 2K_1\, \cos q\,$ 
using an even number of Fourier moments $q \in  \{ \pm \pi/L, \pm 3 \pi/L,\, \cdots\,,
\pm (L-1) \pi/L\}$ \cite{Felderhof}. Thereby, in the limit of $T \to 0$ the dynamics is 
typically diffusive as the inverse gap or relaxation time becomes $\propto L^2$. For 
$J_2 \ne 0$ however, the problem is no longer soluble by analytic treatments but parity 
conservation still holds. Thus, in Sec.\,III we shall restrict the numerical analysis to 
states having only an even number of domain walls, as is natural when PBC are set in 
the original spin system. 

As for the dynamics invariance with respect to the sign of $J_1$ referred to earlier in 
this Section, consider for that matter the global spin rotation
\vskip -0.5cm
\begin{equation}
\vec \sigma \to R\; \vec \sigma \, R^{-1},\;\, R = \exp \left( -i \,\frac{\pi}{2}\,
\sigma^x \right)\,,
\end{equation}
under which $\sigma_j^z \to -\sigma_j^z$ and $\sigma_j^{\pm} \to \sigma_j^{\mp}$. 
Hence, the effect of this  similarity transformation on Eqs.\,(\ref{s-pairing}) and (\ref
{diagonal}) is tantamount, respectively, to the substitutions $B \to -B$ and  $C_{\pm} 
\to - C_{\pm}$, while leaving Eq.\,(\ref{s-diffusion}), $A_{\pm},\,D_{\pm}$, as well as 
the whole spectrum of the evolution operator unaltered. But recalling the definitions of 
$P$ and $Q$ [\,Eq.\,(\ref{pq})\,] entering in these coefficients [\,Eqs.\,(\ref{coef1}) and 
(\ref{coef2})\,], it is then clear that this equivalent description here would just merely 
correspond to the substitution $J_1 \to -J_1$, as previously argued on more intuitive 
grounds.

Let us finally comment that also three-body interactions correlated with single-spin 
flips already appear at the level of the Glauber rates of Eq.\,(\ref{rates}). Specifically, 
exploiting basic relations of the hyperbolic functions, and using the above $C_{\pm}\!$ 
and $D_{\pm}$ coefficients it can be readily shown that those rates actually deploy 
terms of the form
\vskip -0.25cm
\begin{equation}
W \left( S_i \to -S_i \right) = \frac{1}{2} - \frac{S_i}{8} \,\big[\,\big(\,S_{i-1} 
+ S_{i+1} \big) \left(\,C_{_{\!+}} + C_{_{\!-}}\, S_{i-2} \, S_{i+2}\right) - \\
\big(\,S_{i-2} + S_{i+2} \big) \left(\,D_{_{\!+}}+ D_{_{\!-}}\,S_{i-1}\, 
S_{i+1}\right) \big]\,,
\end{equation}
which in turn would yield up to four-body interactions in the original spin evolution 
operator.

\section{Numerical Results}

Armed with the contributions of Eqs.\,(\ref{s-pairing}), (\ref{s-diffusion}), and 
(\ref{diagonal}) operating on a generic kink state $\vert \sigma \rangle$, we next turn 
to the exact evaluation of spectral gaps in finite systems within low temperature 
regimes. As a consistency check, first we verified that transforming the Boltzmann 
distribution $\vert P_B \rangle$ with the above diagonal $\mathbb D$
\begin{equation}
\left\vert \psi_0 \right\rangle = \sqrt{\cal Z\;}\, \mathbb D \,\left\vert P_B \right\rangle 
= \frac{1}{\sqrt{\cal Z\;}}\sum_{\sigma} e^{-\frac{\beta}{2}\,E_{\sigma}}\vert 
\sigma \rangle\,,
\end{equation}
produces in fact, by construction, the ground state of ${\cal H}_{nd}^{^{(pair)}}\!\! 
+ {\cal H}_{nd}^{^{(di\!f\!f)}}\!+ {\cal H}_d$ with eigenvalue $\lambda_0 = 0$ 
(the kink partition function ${\cal Z}$ here acting merely as a normalization factor). 
This also served to start up a Lanczos algorithm \cite{Lanczos} with random initial 
states but chosen orthonormal to that equilibrium $\left\vert \psi_0 \right\rangle$. 
In turn, the  states generated by the Lanczos recursion also were subsequently 
reorthonormalized to $\left\vert \psi_0 \right\rangle$. Thereafter, we obtained the 
first excited eigenmodes of the evolution operator using periodic chains of up to 24 
sites (the main limitation for that being the $2^{L-1}$ dimensions of the kink space). 
Also, as a further preliminary test we retrieved the gap $\lambda_1 (K_1, 0)  = 2 
\left(1 - \tanh 2K_1\, \cos \frac{\pi}{L}\right)$ of the standard Glauber dynamics 
\cite{Felderhof}, and checked out the aforementioned symmetry $\lambda_1 (K_1, 
K_2) = \lambda_1 (-K_1, K_2)$.

Besides length limitations, another restrictive issue to point out here is that for $J_2 
< 0$ the Lanczos convergence slows down progressively as temperature is lowered 
because, as we shall argue and corroborate in a moment, the spacing of low lying 
levels gets arbitrarily small even in finite chains. Due to the smallness of those gaps 
at least quadruple precision is needed, but for $r = -J_2 /\vert J_1 \vert \agt 0.3$ and
temperatures below $T/ \vert J_1 \vert  \sim 0.05$ the pace of convergence becomes 
impractical for the larger sizes at hand. Thus, we shall content ourselves with providing 
results within the weakly frustrated region $0 < r \le 0.3$, where nonetheless a clear 
universal trend already shows up. By contrast, for non-competing interactions the 
spectral gaps of finite systems remain finite even in the limit of $T \to 0$ (alike the 
usual case of $J_2 =0$), and the Lanczos convergence is faster. However, it will turn 
out that there are three types of behavior to examine depending on whether $-1 < r 
< 0,\, r = -1,$ or $r < -1$.

\subsection{Weak competing interactions}

As was referred to in Sec.\,I,  for $0 < r < 1$ the $T=0$ dynamics rapidly leaves the 
system in a metastable state of wandering kinks separated by at least two lattice 
spacings \cite{Redner,Sen}. At low but non-zero temperatures however, we see from 
Table \ref{tab1} that for $0 < r < 1/2$ the energy barrier ($4 \vert J_2 \vert$)
responsible for that restriction corresponds to that of the second and third diffusion 
processes. But now those barriers can be surmounted provided time scales comparable 
to $2/ (1 - \tanh 2 \vert K_2 \vert) \sim e^{\,4\, r K_1 }$ are considered. As a result, all 
ferromagnetic domains then rapidly coalesce just following the energy gradient $-4 ( 
J_1- \vert J_2 \vert ) < 0$ of the first pairing process depicted above. Therefore  it is 
reasonable to expect that even in a finite system the relaxation time to equilibrium 
should diverge as $e^{\,4\, r K_1 }$. 

\begin{figure}[htbp]
\vskip -2.85cm
\hskip -0.75cm
\includegraphics[width=0.7\textwidth]{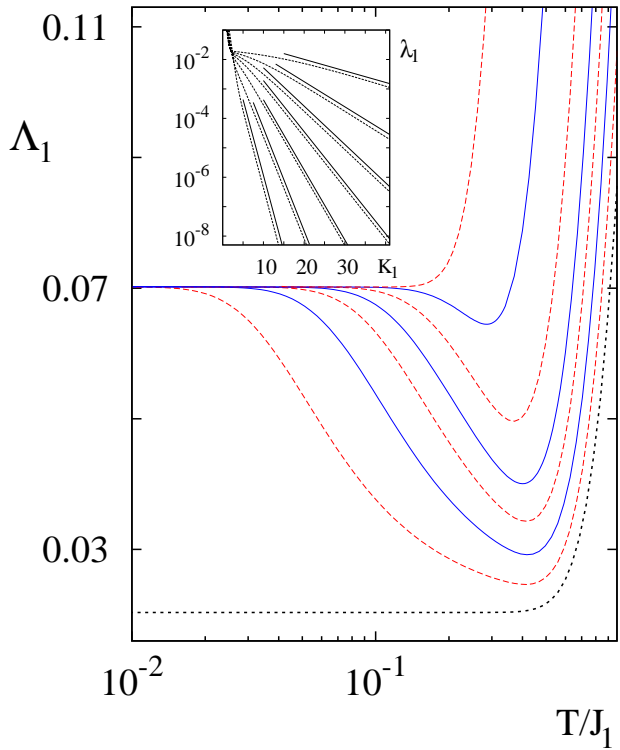}
\vskip -4.5cm
\caption{(Color online) Normalized spectral gaps $\Lambda_1 = e^{\,4\, r K_1 }
\lambda_1$ with frustrated interactions for $L = 22$ on approaching low temperature 
regimes and taking  $r = $ 0.3,  0.2,  0.1375,  0.1,  0.075, 0.05, 0.025 (from top to 
bottom, alternating between dashed and solid lines). All $\Lambda$'s saturate towards 
a common value $\propto L^{-z}$ (see Fig.\,\ref{two}). For comparison, the lowermost 
doted curve denotes the soluble case $r = 0$. The straight solid lines in the inset, 
closely following the asymptotic behavior of the corresponding data, are fitted with 
slopes $-4 \,r$.}
\label{one}
\end{figure}

\begin{figure}[htbp]
\vskip -2.8cm
\hskip -10cm 
\includegraphics[width=0.65\textwidth]{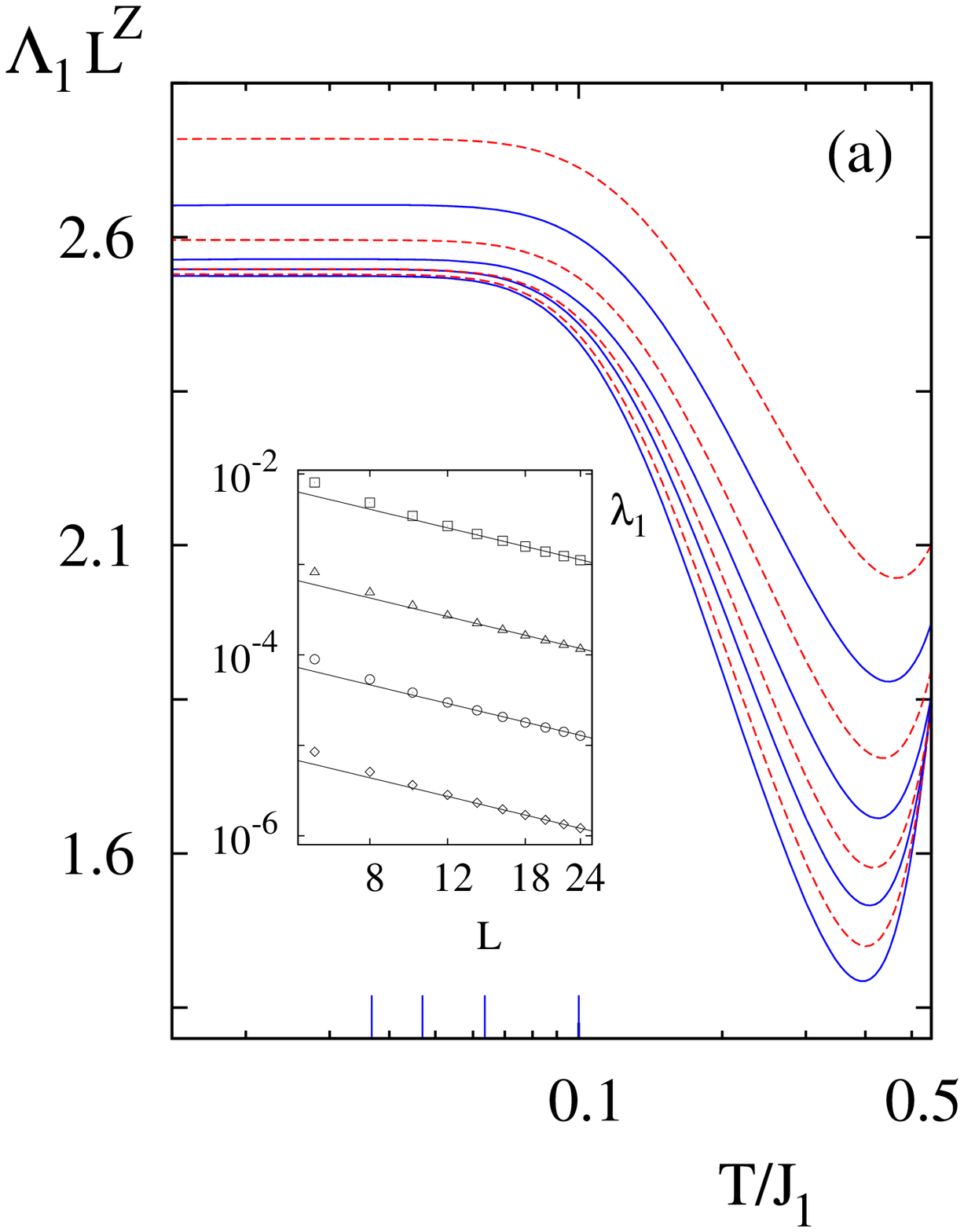}
\vskip -14.9cm
\hskip 6.2cm 
\includegraphics[width=0.65\textwidth]{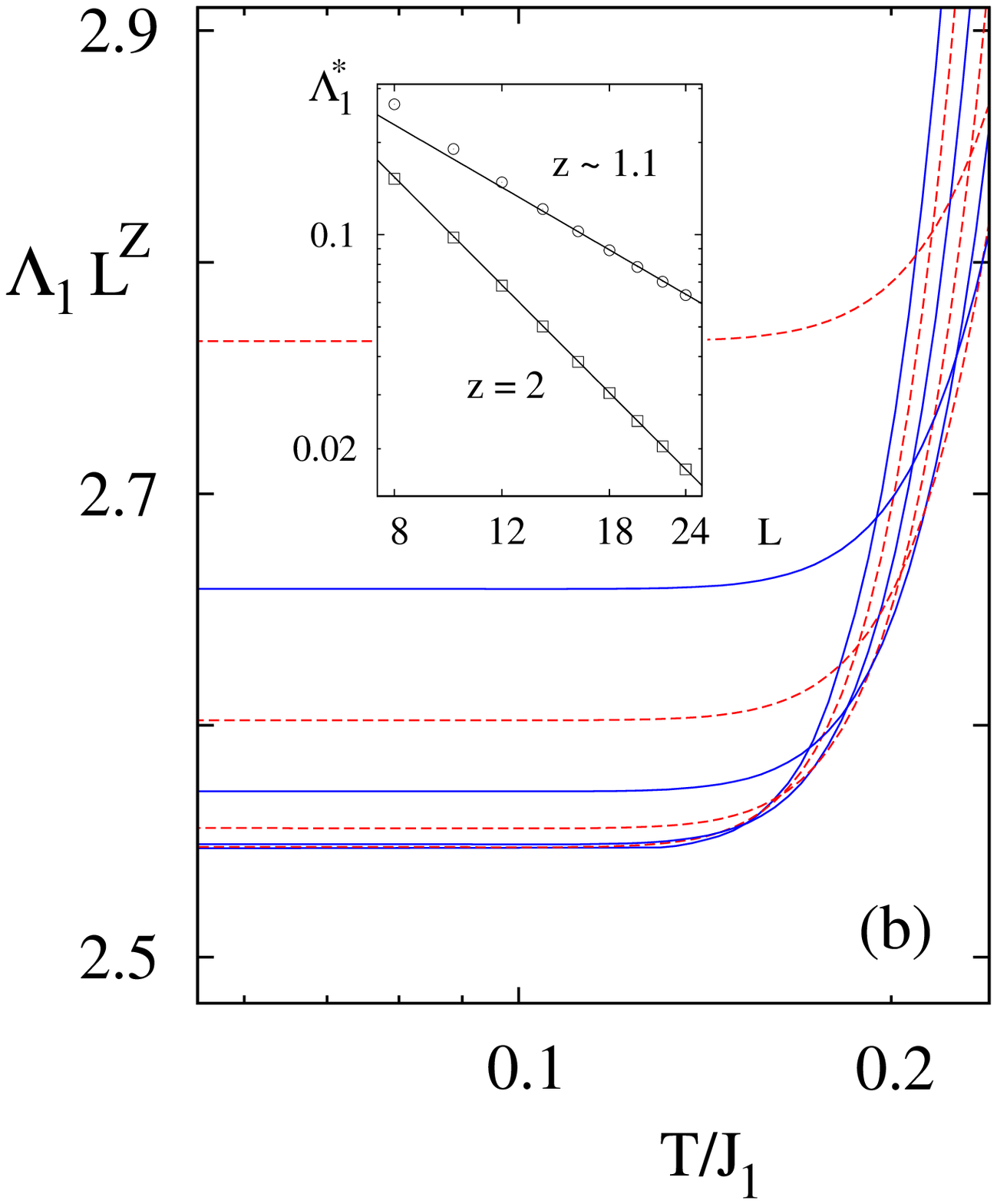}
\vskip -4.75cm
\caption{(Color online) Finite-size scaling of normalized spectral gaps $\Lambda_1 = 
e^{\,4\, r K_1 } \lambda_1$ for (a) $r = 0.1$, and (b) $r = 0.3$. In the main panels, 
sizes increase downwards as $L = 2 k$ with $5 \le k \le 12$ alternating between dashed 
and solid lines. At low temperatures the data collapse of larger sizes was attained upon 
setting a common dynamic exponent $z \sim 1.1$ (however, see extrapolations of Sec. 
III\,C). The inset of (a) shows the plain spectral gap behavior for the temperatures 
indicated by vertical lines below, always consistent with an $\sim L^{-1.1}$decrease. 
Similarly, the inset of (b) exhibits the typical finite-size decay of $\Lambda_1$ at 
saturation [\,Eq.\,(\ref{saturation})\,] for all $r$'s considered in Fig.\,\ref{one}. For 
comparison, the diffusive case of $J_2 =0$ is also shown here.}
\label{two}
\vskip 0.3cm
\end{figure}

In Fig.\,\ref{one} we test this hypothesis for several values of $r \in (0,\,0.3]$ as 
temperature is lowered. The saturation trends of the `normalized' gaps $\Lambda_1 
\equiv e^{\,4\, r K_1 }\, \lambda_1$ evidently confirm this expectation in this weakly 
frustrated region. This is also evidenced by the $-4\, r$ slopes  shown in the inset
which closely follow those plain gap decays. Note that even a slight deviation from 
these slope values would result in strong departures from the saturation regimes
exhibited in the main panel. In that temperature limit  each $\Lambda_1$ appears to 
be independent of $r$, although they all differ from the usual Glauber gap $\simeq
\pi^2 /L^2$ of finite systems at $T= 0$, thus already signaling a discontinuous
behavior at $r=0$ (see also Sec.\,III\,B). In fact, and more importantly, there the
dynamic exponents are no longer diffusive, as indicated in Fig.\,\ref{two} by the data 
collapse towards larger sizes. This was attained upon choosing an exponent $z \sim 
1.1$, in turn consistent with the slopes read off from the inset of Fig.\,\ref{two}a where 
the finite-size behavior of the plain gap is exhibited at various temperatures. Analogous 
results were obtained for other weakly competing ratios, always recovering similar
$z$-\,values, and as mentioned above, yielding only size dependent $\Lambda_1$'s 
at saturation regimes. It seems then natural to put forward the hypothesis that in the 
limit of $T \to 0^+$ and large $L$ these normalized gaps should all scale as
\begin{equation}
\label{saturation}
\Lambda_1^{\!^*} (L) \equiv \lim_{T \to 0^+} e^{-\,4 K_2 }\,\lambda_1 (K_1,K_2,L) 
\propto L^{-z},
\end{equation}
the proportionality factor just being a constant of the order of $\sim 2.55$, as observed 
in the main panels of Fig.\ref{two}.

In the inset of Fig.\,\ref{two}b we check this out for several values of $r \in (0,\,0.3]$,
in all cases using temperatures where the saturation trends were already settled. There, 
the height of each data symbol encapsulates the results obtained for every $r$ in 
that range. In fact this latter could be extended up to $r = 0.4$ although due to the 
convergence slowing mentioned before, no results are in our disposal for $L \ge 22$. 
Yet, the key issue here is that no matter how weak the frustration might be, our
analysis yields a dynamic exponent far appart from the diffusive $z =2$ obtained for 
$r=0$ (also shown for comparison). Let us anticipate that the nearly ballistic exponent 
($z \simeq 1$) to arise in Sec.\,III\,C from extrapolations to the thermodynamic limit, 
also enables us to think of the large amplitudes of Eq.\,(\ref{saturation}) as small 
asymptotic velocities of growth. Specifically, recalling that at times of order $1/ 
\lambda_1$ both the average domain size and correlation length become comparable 
to the system size, clearly from Eq.\,(\ref{saturation}) it then follows that the ordering 
scale must coarsen almost linearly in time with an arbitrarily slow velocity $\propto  
e^{\,-4\, r K_1 }$.

It would be desirable to complete the weakly frustrated picture in the region $1/2 < r < 
1$ but there, once more, the smallness of spectral gaps precluded convergence in the 
larger chains (here increasing by four sites to match the ground state periodicity). As 
before, the gap turned out to decrease as $e^{\,-4\, r K_1 }$ though as temperature 
was lowered  the relenting convergence pace impeded us to obtain further results for 
$L \ge 20$.

\subsection{Non-frustrated regime}

\begin{figure}[htbp]
\vskip -2.35cm
\hskip -1.2cm
\includegraphics[width=0.68\textwidth]{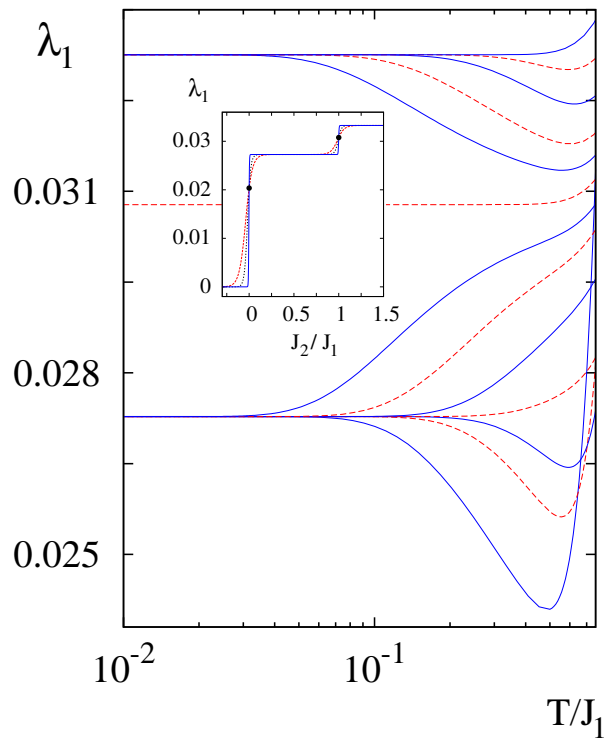}
\vskip -4.85cm
\caption{(Color online) Gaps of non-competing chains with $L = 22$. From top to 
bottom the main panel depicts the branches corresponding respectively to (i) $-r =
J_2 / \vert J_1 \vert > 1$\, (2, 1.35, 1.2, 1.1, 1.05), (ii) $r = -1,$ and (iii) $0 < -r < 1$\, 
(0.95, 0.9, 0.8, 0.6, 0.3, 0.2, 0.1), alternating between solid and dashed lines. Alike 
the frustrated regime (Fig.\,\ref{one}), at low temperatures each branch saturates 
towards common values $\propto L^{-z}$ (see Fig.\,\ref{four}). The inset exhibits the 
discontinuous tendency of these latter as temperature is lowered ($T/J_1 = 0.15, 0.07, 
0.01$). Filled circles stand at the intersection of cases $r = 0$, and $r= -1$. Gaps at the 
leftmost regime (weakly competing, $r > 0$) decay as $e^{\,-4\, r K_1 }$ (see inset of 
Fig.\,\ref{one}).}
\label{three}
\end{figure}

Next, we move on to the case of non-competing interactions, i.e $J_2 > 0$, for which 
there are no longer energy barriers hindering the access to low temperature 
equilibrium. Thus, the relaxation times of finite systems remain bounded even at
$T=0$. However, in that temperature limit it can be seen that a discontinuity already 
appears at the level of transition rates both at $r = 0$ and $r = - 1$ . In the first case 
this affects either the second or third diffusion process of Table~\ref{tab1}, whereas 
for $r = - 1$ it appears either in the first or fourth pairing process quoted there. 
Correspondingly, these discontinuities also affect the diagonal elements of the 
evolution operator via the $C_{\pm}$ and $D_{\pm}$ coefficients defined in 
Eq.\,(\ref{coef2}), and are ultimately reflected in the low temperature gaps shown 
in Fig.\,\ref{three}. Its main panel clearly distinguishes different regimes below, at, 
and above $r = - 1$, each one encompassing a branch of spectral gaps with a common 
saturation point in the low temperature limit. The inset of Fig.\,\ref{three} displays 
the stepwise trend of this latter for both non-competing and competing ratios, while a
similar step pattern is followed by other low lying $\lambda$-levels (not shown). Note 
here that in nearing $T = 0^+$ the left discontinuity at $J_2 =0$ is signaling the 
abrupt crossover from the non-diffusive kinetics studied in Sec.\,III\,A to the usual
Glauber dynamics with no energy barriers. 

\begin{figure}[htbp]
\vskip 0.1cm
\hskip -13cm
\includegraphics[width=0.35\textwidth]{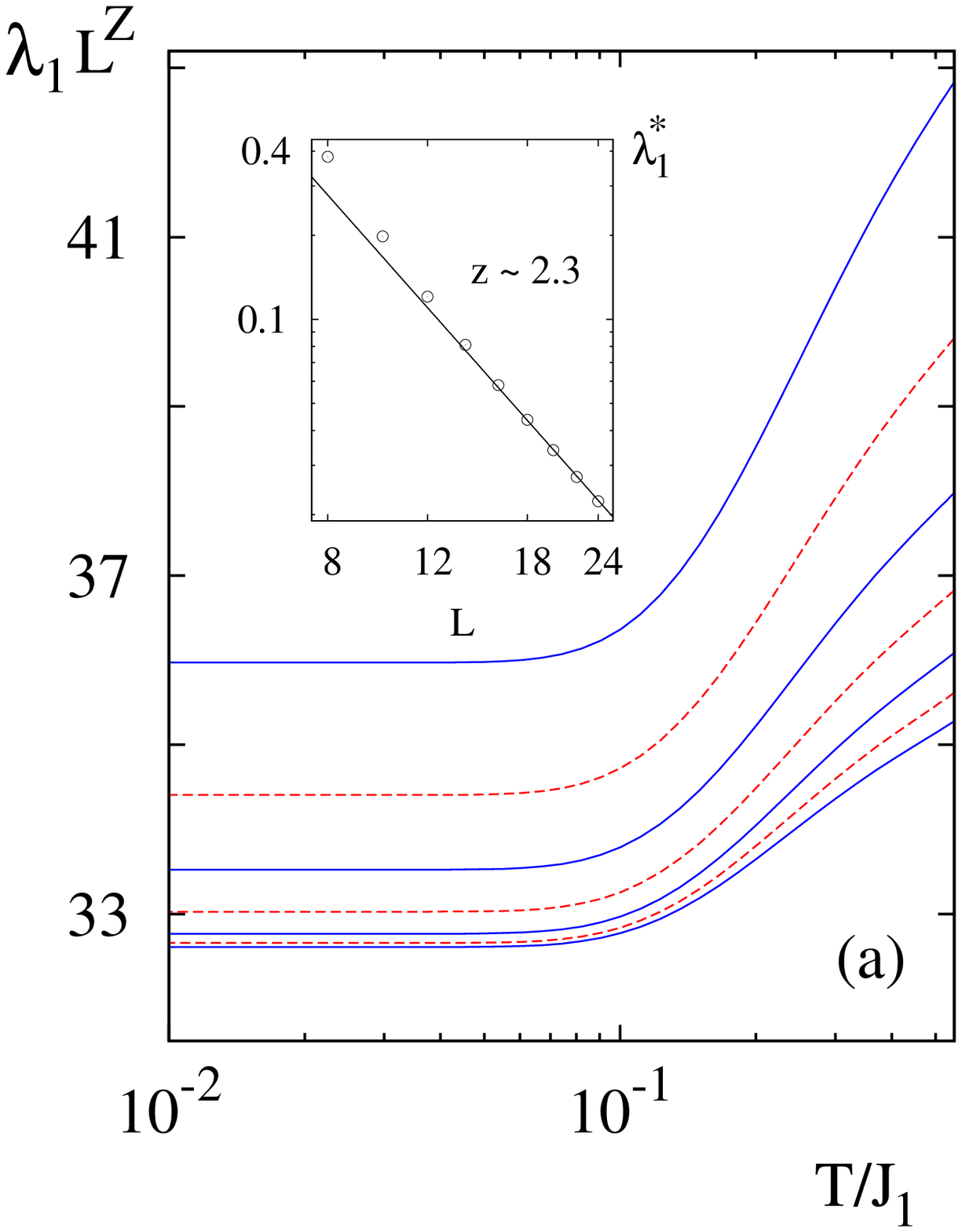}
\vskip -6.72cm
\hskip -0.71cm
\includegraphics[width=0.35\textwidth]{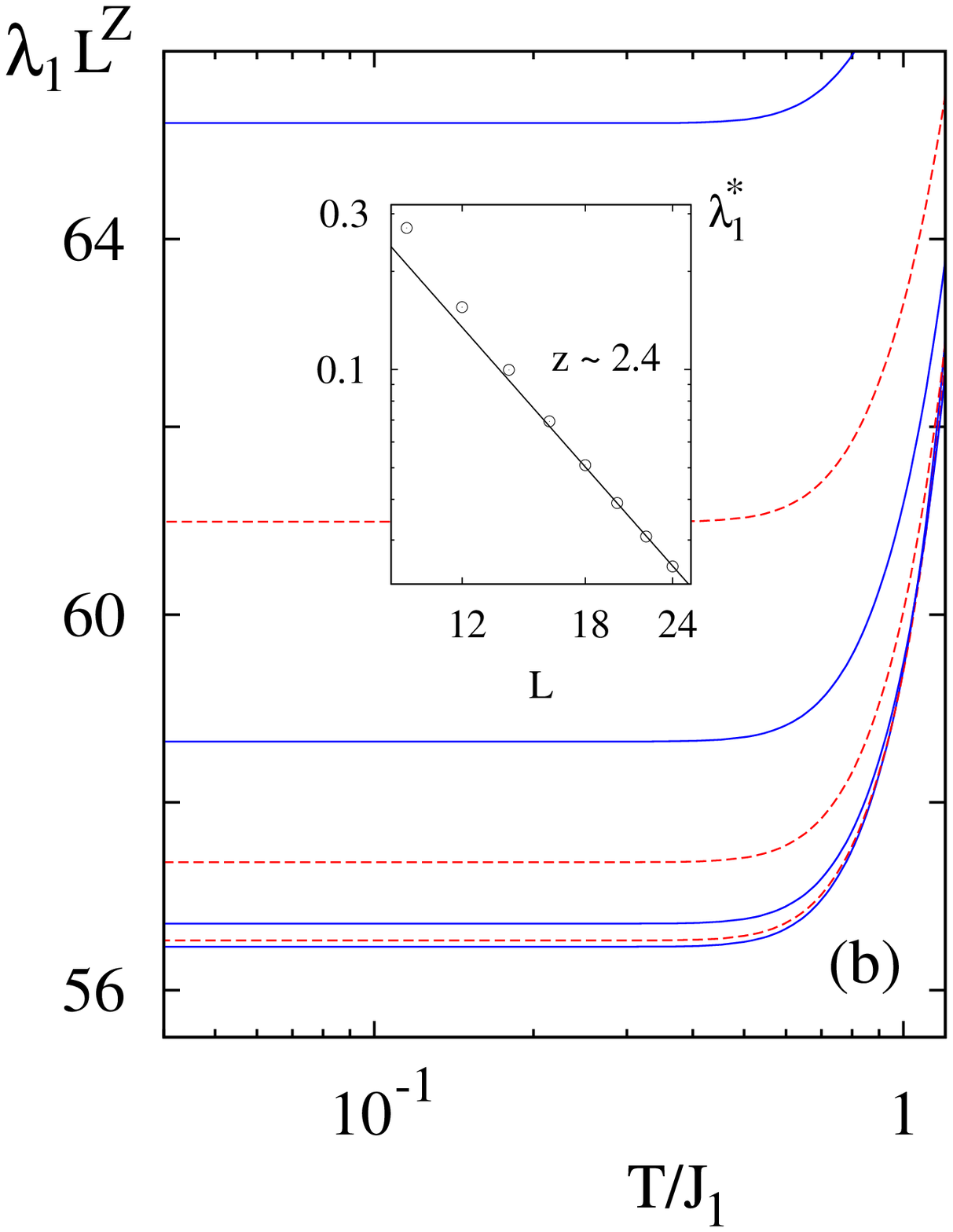}
\vskip -6.4cm
\hskip 11.6cm
\includegraphics[width=0.35\textwidth]{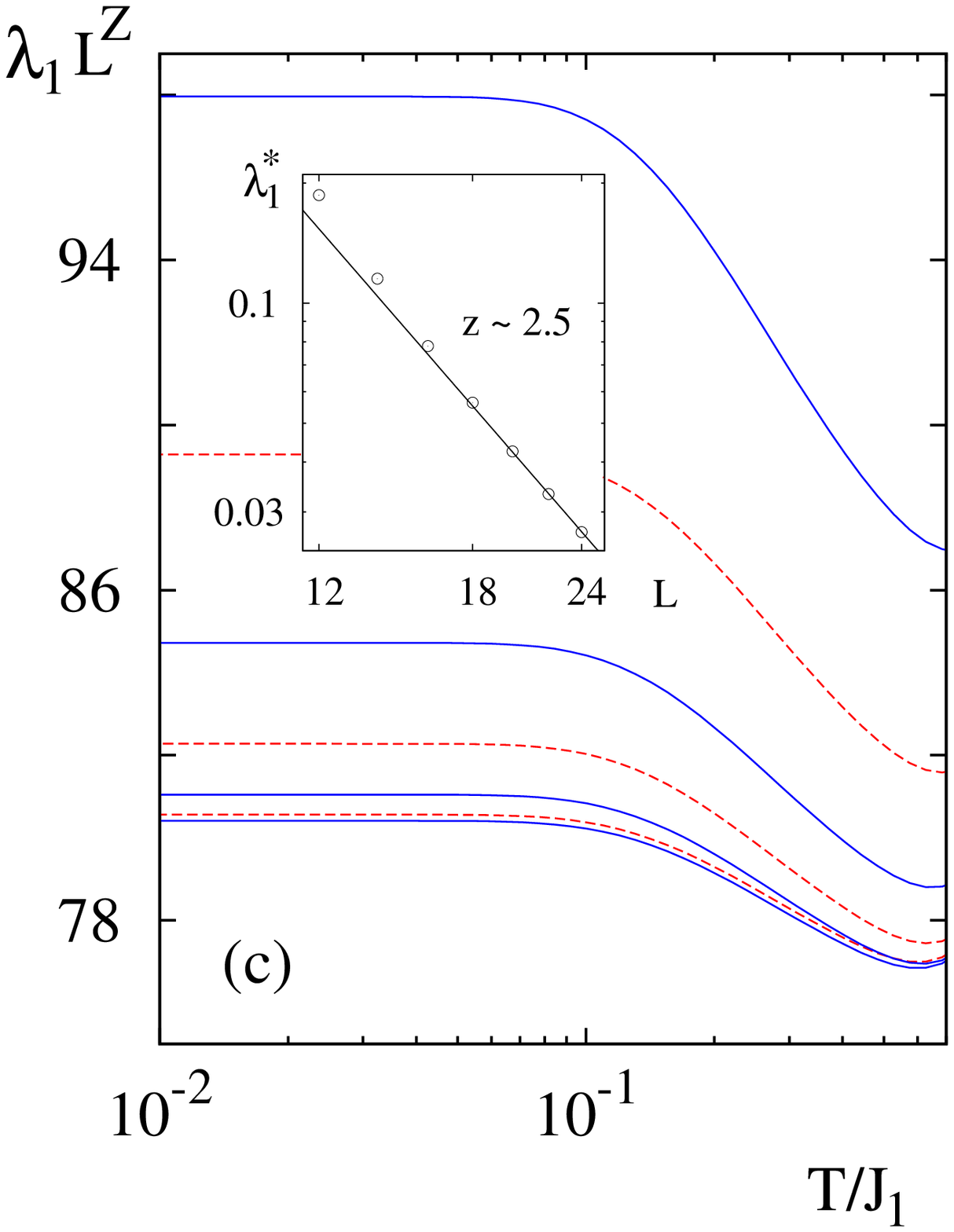}
\vskip 0.4cm
\caption{(Color online) Typical finite-size scaling of gaps for non-frustrated coupling
regimes (Fig.\,\ref{three}), taking (a) $J_2/\vert J_1\vert = 0.9$, (b) 1, and (c) 1.1. 
Lengths, alternating between solid and dashed lines, increase downwards as $L = 2k$ 
with $6 \le k \le 12$. At low temperatures the data collapse of larger sizes in each 
branch was obtained upon choosing $z \sim$ 2.3, 2.4 and 2.5 respectively. In turn, 
these values are used as slopes to fit the typical finite-size decay of these gaps at 
saturation (denoted by $\lambda_1^{\!^*}$ in the insets). See however 
extrapolations of  Sec. III\,C.}
\label{four}
\end{figure}

To characterize the non-frustrated dynamics at large times, as before, we resort to a 
finite-size scaling analysis of these gaps in each of the above regimes. This we do in 
the main panels of Fig.\,\ref{four} which exhibit the data collapse onto larger sizes 
obtained at and near $r = -1$. As might be expected from Fig.\,\ref{three}, the
dynamic exponents producing these collapses turn out to be branch dependent only.
In each case their values, all subdiffusive, can also be read off from the slopes that fit 
the finite-size decay of the corresponding low temperature gaps, now converging much 
faster than those of the case $r > 0$. This is displayed in the insets of Fig.\,\ref{four} 
where each data point is also representative of all $r$'s studied in the saturation
regimes of Fig.\,\ref{three}. In particular, the gaps shown in the last inset are also
typical of several coupling ratios which we studied additionally in the region $-r \gg 1$.
However, the proximity of the resulting exponents calls into question whether these 
actually stem from different asymptotic dynamics per branch or are rather merely 
products of our finite-size limitations. In that latter respect, and aiming to observe in 
more detail those size effects, we now turn to standard extrapolations \cite{Henkel, 
Guttmann} which go a step further than the collapse trends of Figs.\,\ref{two} and 
\ref{four}.

\subsection{Extrapolations}

\begin{figure}[htbp]
\includegraphics[width=0.485\textwidth]{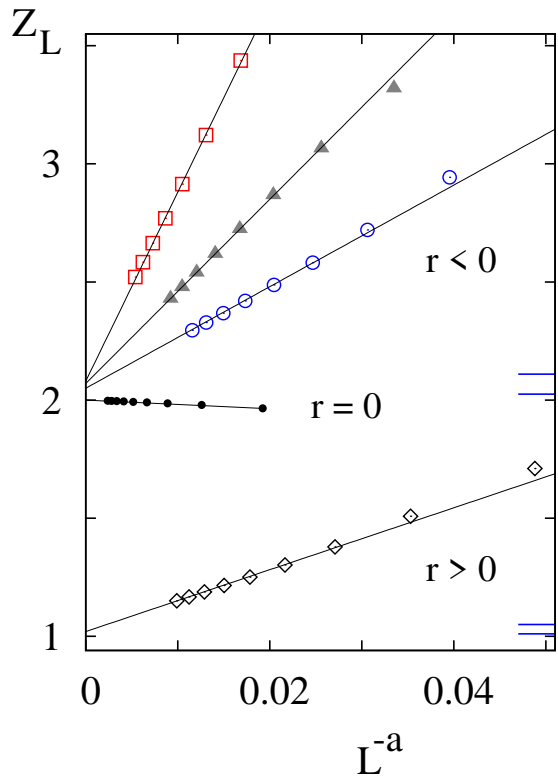}
\vskip 0.75cm
\caption{(Color online) Extrapolations of effective dynamic exponents  
[\,Eq.\,(\ref{sequence})\,] within the regions $0 < r = -J_2/\vert J_1\vert  \le 0.3$
(rhomboids), $r = 0$ (dots), $0 < -r < 1$ (circles), $r = -1$ (triangles), and $r < -1$ 
(squares). Non-linear fittings in the weak frustrated regime $(a \sim 1.45)$ yield a 
dynamic exponent $z \simeq 1.03 \pm 0.02$, while those for circles $(a \sim 1.40)$, 
triangles $(a \sim 1.47)$, and squares $(a \sim 1.64)$ yield slightly subdiffusive 
exponents within the confidence interval $[2.02 \,, 2.11]$. This latter and that of 
the weak competing situation are depicted by rightmost horizontal lines. The usual $r 
= 0$ case converges faster $(a \sim 1.95)$ towards the exact exponent within the 
same range of sizes.}
\label{five}
\end{figure}

Having evaluated the spectral gaps of the evolution operator on finite-sizes, an
improved estimation of its dynamic exponents along with a measure of their 
convergence can be obtained upon considering the sequence of effective 
approximants defined as
\vskip -0.4cm
\begin{equation}
\label{sequence}
Z_L = \frac{\ln[\,\lambda_1^{\!^*}(L) / \lambda_1^{\!^*}(L-2) \,]}{\ln[\,(L-2)/L\,]}\,.
\end{equation}
These are simply the local slopes of the insets in Figs.\,\ref{two} and \ref{four}, thus  
representing successive estimates of the gap closing in the low temperature limit. 
Here, all $L$-lengths are even so as to allow for the $J_1\! \to\! -J_1$ symmetry referred 
to above. Near criticality ($T\! = 0^+$) this sequence of approximants is generally 
assumed to converge logarithmically \cite{Henkel,Guttmann} as 
\begin{equation}
Z_L = z + \alpha_1 L^{-a_1} +  \alpha_2 L^{-a_2} + \cdots\,,
\end{equation}
with $\alpha$- constants and $a$- exponents taken such that $0 < a_1 < a_2 < \cdots$ 
To minimize the number of fitting quantities we keep only the leading order term of this 
expansion which just leave us with a nonlinear least squares fit of three parameters. 
The results of this are depicted in Fig.\,\ref{five} which summarizes the trends of these 
sequences across the situations considered in Secs.\,III\,A and III\,B. As anticipated, for 
weak competing interactions the extrapolated exponents thus resulting from these 
nonlinear regressions turn out to be almost ballistic ($z =1$), while those arising from 
the non-competing regimes become slightly subdiffusive ($z > 2$), specifically
\begin{equation}
z \simeq 
\begin{cases}
1.03 \pm 0.02\,,\;\;{\rm for}\,\;\; 0 < r \alt 0.3\,,
\vspace{0.1cm} \cr
2.05 \pm 0.03\,,\;\;{\rm for}\,\;\; 0 < -r  < 1\,,
\vspace{0.1cm} \cr
2.07 \pm 0.04 \,,\;\;{\rm for}\,\;\;  r = -1\,,
\vspace{0.1cm} \cr
2.08 \pm 0.03\,,\;\;{\rm for}\,\;\;  r < -1\,.
\end{cases}
\end{equation}
In each case the degree of convergence of the corresponding sequences is similar (see 
$a$- exponents mentioned in Fig.\,\ref{five}) but slower than that obtained in the 
standard dynamics of $r = 0$ (also shown for comparison), and whose extrapolated 
$z$- exponent  $\simeq 1.997 \pm 0.004$ comes out pretty close to its actual value.

We also fitted these sets of finite-size results using sequence transformation methods 
such as Vanden Broeck-Schwartz type ones \cite{VBS}, which yield basically analogous 
$z$'s within the above margins of error. However, since in practice it is never really
clear whether the assumed asymptotic behavior is sufficiently well realized by the data 
available \cite{Henkel}, the slight differences among these subdiffusive exponents 
might well be ascribed to our finite-size limitations. In that sense, the merging of
confidence intervals (rightmost center of Fig.\,\ref{five}) suggests a common
characterization of the three non-competing branches of Fig.\,\ref{three} within 
an error margin of $\sim 4 \%$. On the other hand, for $J_2 \gg \vert J_1 \vert$ one 
should recover the usual diffusive dynamics as the system would then reduce to two 
independent Glauber processes, thus suggesting a slight overestimation of dynamic 
exponents in this region. (Note that this case is essentially different from the regime 
$J_2 \to 0^-$, because for that latter all kinks separated by two dual spacings yet 
strongly repel each other in the low temperature limit).
\vskip -0.3cm

\vspace{-0.25cm}

\section{Concluding Discussion}

Recapitulating, we have constructed a symmetric spin-$\frac{1}{2}$ representation 
of  the 1D Glauber dynamics with both first and second neighbor interactions by which 
we analyzed the scaling behavior of relaxation times of finite chains under various 
situations at infinitesimal temperature. The resulting time scales were read off from 
the spectral gaps of the corresponding evolution operator built up in Sec.II [\,Eqs.\,(\ref
{s-pairing}), (\ref{s-diffusion}), and (\ref{diagonal})\,] to account both for the 
correlated kink pairing and diffusion processes schematized in Table~\ref{tab1}. 

Special attention was paid to weakly frustrated regimes ($J_2 < 0$) where energy 
barriers hindering the coalescence of magnetic domains (second and third diffusion 
processes of Table~\ref{tab1}) cause the relaxation time to grow unbounded as 
$e^{\,4\, \vert K_2 \vert}$, even for finite systems (Fig.\,\ref{one}). However a clear 
finite-size scaling regime [\,Eq.\,(\ref{saturation})\,] turned out to take over the low but 
non-zero temperature limit of those time scales, at least within our accessible range of 
competing interactions (main panels of Fig.\,\ref{two} and inset of Fig.\,\ref{two}b). 
In stark contrast to the non-coarsening scenario at $T=0$ \cite{Redner, Sen}, that led 
us to suggest a nearly ballistic dynamic exponent $z \simeq 1.03(2)$ (lowermost 
sequence extrapolations of Fig.\,\ref{five}), with which domains coarsen almost linearly 
in time. Still, owing to the mentioned coalescence barriers, they do so at arbitrarily slow 
velocities $\propto e^{\,-4\, \vert K_2 \vert}$ as suggested by Eq.\,(\ref{saturation}). 
It would be important to examine also the weakly frustrated region $1/2 <  -J_2/ \vert 
J_1\vert  < 1$ where, to reach one of its four-fold degenerate states ($\cdots \vert \bullet 
\bullet\, \vert \circ \circ \,\vert \cdots$), the dynamics must first create kink pairs so as 
to break up isolated ferro domains of length $\ge 3$ \cite{Redner}. However, as was 
referred to in Sec.\,III\,A, in lowering the temperature the progressive smallness of 
spectral gaps precluded the Lanczos convergence on larger chains; a drawback which 
already appeared for $-J_2/\vert J_1\vert \agt 0.3$, and still remains an open issue.
On the other hand, analytical approximations in that region would have to arbitrarily 
decouple the infinite hierarchy of equations of motion embodied in the evolution 
operator. But in view of the discontinuous scaling behavior obtained even for $J_2 \to 
0^-$,  such procedure might well result inappropriate at low temperature regimes.

When it comes to non-competing regimes ($J_2 > 0$), the convergence problem no 
longer emerges as there are no metastable states slowing down the dynamics anymore, 
thus relaxation times of finite systems are kept bounded even for $T = 0$. This 
facilitated the numerical analysis of spectral gap discontinuities occurring as a result of 
those that naturally appear in the limit of $T \to 0$ on the transition probability rates at 
$J_2/\vert J_1\vert = 0$ and 1 (Fig.\,\ref{three}). In turn, that resulted in three branches 
of gaps whose typical scaling regimes were initially characterized by somewhat similar 
dynamic exponents (Fig.\,\ref{four}). The proximity among these latter however 
became much closer after extrapolating the sequence of finite-size effective 
approximants [\,Eq.(\ref{sequence})\,] summarized in Fig.\,\ref{five}, and to the point 
of suggesting an almost diffusive exponent $\forall\, J_2 > 0$.

Finally, and with regard to a possible extension of this study, it would be interesting to 
apply these scaling techniques also to the 1D Kawasaki dynamics \cite{Puri} (say for 
$J_1 > 0$) where  due to the strict constancy of magnetization, initial disordered states 
quenched to zero temperature get stuck in metastable phases already for $J_2 =0$ 
\cite{Spirin}. As before, these just involve ferromagnetic domains of two or more spins, 
the number of  such configurations growing as $[\,(1 + \sqrt 5)/2]^L$. However, for 
$1/2 < J_2 /J_1 < 1$ it can be readily verified that states having two or more consecutive 
domains of length two would be excluded from that metastable set. That further 
constraint would still leaves us with a number of configurations growing as $\gamma^L$, 
though with $\gamma$ actually {\it smaller} than the golden ratio. For infinitesimal 
temperatures this raises the question of whether such reduction of metastability could
also bring about strong changes in the scaling regimes of this phase separation 
dynamics. Further work in that direction is in progress. As for $d > 1$, in principle 
a quantum spin representation could also be constructed though its Lanczos 
diagonalization would be restricted to rather few and small lattices, thus rendering 
finite-size scaling impractical.

\section*{Acknowledgments}

I thank H. D. Rosales and F. A. Schaposnik for helpful discussions. Support from 
CONICET and ANPCyT, Argentina, under Grants No. PIP 2012--0747 and No. PICT 
2012--1724, is acknowledged.




\begin{thebibliography}{99}
\bibitem{note1} This latter is immaterial as it does not affect the relative spin 
orientation between second neighbors in the ground state.
\end{thebibliography}
\end{document}